\documentclass[aps,reprint,amsmath,amssymb,prb,superscriptaddress,floatfix,pdftex]{revtex4-2}
\usepackage[pdftex]{graphicx}
\usepackage{dcolumn}
\usepackage{bm}
\usepackage{braket}
\usepackage{color}

\allowdisplaybreaks[4] 

\begin{document}


\title{Theory of Rashba splitting in quantum-well states}



\author{Mitsuaki Kawamura}
\email{mkawamura@ds.itc.u-tokyo.ac.jp}
\affiliation{Information Technology Center, 
  The University of Tokyo, Bunkyo 113-8658, Japan}

\author{Taisuke Ozaki}
\affiliation{Institute for Solid State Physics, 
  The University of Tokyo, Kashiwa 277-8581, Japan}


\date{\today}

\begin{abstract}
  We present a theory pertaining to the asymptotic behavior of Rashba energy splitting in a quantum-well state (QWS).
  First, unlike previous studies, we derive $\textbf{k}$-linear Rashba term from a first-principles Hamiltonian in a physically convincing manner.
  The $\textbf{k}$-dependent in-plane intrinsic magnetic-field term originates from the spin--orbit interaction and hybridized $s$-$p_z$ orbital, whereas a steep nucleus potential realizes the linearity for the $\textbf{k}$ of the effective magnetic field.
  Next, we analyze the Rashba effect of a QWS using a one-dimensional tight-binding model developed based on the bottom-up approach that is aforementioned.
  The Rashba-splitting behavior of this system is captured from the density at the interface. 
  The density can be expressed analytically as a function of the monolayer number and well depth.
  Finally, we apply our formula to the QWS of a few-monolayers Ag on an Au(111) surface to validate the theory based on a realistic system.
  Our tight-binding analysis qualitatively fits the first-principles result using only two fitting parameters and predicts the optimal condition for achieving a large Rashba splitting.
\end{abstract}


\maketitle

\section{Introduction}

In a system featuring significant spin--orbit coupling (SOC) and broken inversion symmetry, the spin degeneracy can be resolved using the Bloch wavenumber ($\textbf{k}$)-dependent intrinsic Zeeman-like Hamiltonian as follows:
\begin{align}
  \hat{H}_\textbf{k}^\textrm{R} = \alpha_\textrm{R} (\textbf{e}_z \times \textbf{k})\cdot\hat{s}.
\end{align}
This phenomenon is known as the Rashba effect~\cite{WOS:A1984SY04100009}, which was first directly observed on an Au(111) surface via angle-resolved photoelectron spectroscopy (ARPES)~\cite{PhysRevLett.77.3419}.
After this discovery, the Rashba effect was observed on other metallic surfaces, such as Bi~\cite{PhysRevLett.93.046403}, Ag, and Cu~\cite{PhysRevB.98.041404}.
This effect can be applied to a spin-current source and detector in spintronics devices~\cite{sanchez2013spin}.
One of the leading research topics pertaining to Rashba-splitting materials is improving splitting to realize a higher spin--charge current conversion ratio.
A long-range ordered $(\sqrt{3}\times\sqrt{3})R$30\textdegree surface alloy comprising $1/3$ of a monolayer (ML) Bi on an Ag(111) surface significantly enhances this effect~\cite{PhysRevLett.98.186807}, and a bulk BiTeI~\cite{WOS:000291969500017} has a larger splitting.
Nanostructure systems are another exciting platform to observe the Rashba effect.
In InGaAs/InAlAs~\cite{PhysRevLett.78.1335} or InGaAs/InP~\cite{PhysRevB.55.R1958} interfaces, we can modulate the magnitude of energy splitting by adjusting the gate voltage.
Similarly, the layer number of the quantum-well state (QWS) system is a tunable parameter that governs the Rashba effect~\cite{PhysRevB.70.193412,PhysRevB.73.195413,PhysRevB.104.L180409}.
Moreover, in the QWS system, another degree of freedom, namely, multiple localization modes, appears depending on the thickness of the confinement area~\cite{PhysRevB.84.075412,CHIANG2000181}.
Therefore, we can widely and flexibly adjust Rashba energy splitting in the QWS system. 

To design a material system by leveraging the Rashba effect, theoretical and computational schemes are effective as they allow the consideration of synthesized and hypothetical materials.
First-principles calculations based on density functional theory (DFT)~\cite{PhysRev.136.B864,PhysRev.140.A1133}, including SOC~\cite{AHMacDonald_1980}, can well reproduce the spin-splitting energy band of Rashba systems such as the Au(111) surface state~\cite{BIHLMAYER20063888} and Bi surface alloy~\cite{PhysRevB.75.195414,YAMAGUCHI2017688}.
Although the energy band from DFT may not necessarily correspond to the ARPES spectrum, the extremely weak electronic correlation of these materials enables a quantitative comparison of the theoretical and experimental results.
The QWS system of few-monolayers Ag on Au(111) surfaces is an example of such a weakly correlated Rashba system, for which we reproduced the energy band and Rashba-splitting parameter quantitatively in a previous study~\cite{PhysRevB.104.L180409}.
Based on further analysis, we revealed that the envelope function of a QWS at the Ag/Au boundary offers a significant contribution to the energy splitting.
Our theoretical background is based on the result of an earlier study~\cite{Nagano_2009}, which derives the $\textbf{k}$-dependent energy splitting $\Delta \varepsilon_\textbf{k}^{\textrm{SOC}}$ as
\begin{align}
  \Delta \varepsilon_\textbf{k}^{\textrm{SOC}} = 
  |\textbf{k}|\int d^3 r 2\frac{\partial V}{\partial z} |\psi_\textbf{k}(\textbf{r})|^2
\end{align}
by assuming the two-dimensional free electronic wavefunction $\psi_\textbf{k}(\textbf{r})=e^{i{k_x x + k_y y}}u(z)$.
However, as the wavefunction of a realistic system, such as surface alloy and the QWS, is different from that of two-dimensional free electronic state, the applicability of the earlier study is unclear, and we must address the discrepancy between the atomic SOC and the Rashba interaction term.
Additionally, if we can formulate such a bottom-up derivation, then a precise theoretical model for capturing the wide-ranging behavior of the QWS system can be derived.

This study aims to systematically derive a $\textbf{k}$-dependent Rashba splitting formulation for a realistic system.
Additionally, we derive the minimum model for the Rashba effect of the QWS to understand its nature and predict the optimal condition to achieve a large Rashba splitting.
This paper is organized as follows:
The derivation of Rashba energy splitting and a simplified model of the QWS along with its analytical solution are presented in Sec.~\ref{sec_theory}.
In Sec.~\ref{sec_example}, an example of Rashba splitting in a QWS, namely few-monolayer Ag on an Au (111) surface, is presented.
Additionally, a comparison between our analytical result and the result from first-principles calculation is presented.
Finally, we present our summary in Sec.~\ref{sec_summary}.

\section{Theory} \label{sec_theory}

In this section, we explain the basic theory of the Rashba effect in a surface system and then derive a minimum model for the QWS system.
Finally, the model is solved analytically using Green's function.

\subsection{SOC and Rashba splitting}

The SOC term in a Hamiltonian is 
\begin{align}
  \hat{H}^{\textrm{SOC}}_{\sigma \sigma'} 
  = 2 \{\nabla V({\textbf{r}}) \times \hat{\textbf{p}}\} \cdot \textbf{s}_{\sigma \sigma'},
\end{align}
where $\textbf{s}_{\sigma \sigma'}$ are the Pauli matrices times $1/2$, $\sigma (\sigma') = \{\uparrow, \downarrow\}$, $V({\textbf{r}})$ is the nuclear potential, and $\hat{\textbf{p}}\equiv -i\nabla$ is the momentum operator.
The contribution of the Hamiltonian above to the orbital eigenenergy is computed using spinor eigenfunctions $(\psi^{\pm}_{\textbf{k} \uparrow}, \psi^{\pm}_{\textbf{k}\downarrow})$ at wavenumber $\textbf{k}$ and band $n$ as follows:
\begin{align}
  \varepsilon_{\textbf{k}\pm}^{\textrm{SOC}} = 2 \sum_{\sigma \sigma'} 
  \int d^3 r \psi^{\pm *}_{\textbf{k}\sigma}(\textbf{r})
  \{\nabla V({\textbf{r}}) \times \hat{\textbf{p}}\} \cdot \textbf{s}_{\sigma \sigma'} 
    \psi^{\pm}_{\textbf{k}\sigma'}(\textbf{r}).
    \label{eq_soc_eigenenergy}
\end{align}
Here, we consider the energies of two states, indexed as $+$ and $-$, separately.
By assuming the absence of the space dependence of the spin in a single orbital, the Bloch spinor wavefunction $\psi^{\pm}_{\textbf{k}\sigma}$ is written as
\begin{align}
  \left(\begin{matrix}
    \psi^{\pm}_{\textbf{k}\uparrow}(\textbf{r}) \\
    \psi^{\pm}_{\textbf{k}\downarrow}(\textbf{r}) \\
  \end{matrix}\right)
  =e^{i\textbf{k}\cdot\textbf{r}}
  u_{\textbf{k}}(\textbf{r})
  \left(\begin{matrix}
    \chi^{\pm}_{\textbf{k}\uparrow} \\
    \chi^{\pm}_{\textbf{k}\downarrow} \\
  \end{matrix}\right),
  \label{eq_bloch_wavefunction}
\end{align}
where $\chi^{\pm}_{\textbf{k}\sigma}$ is the space-independent spin portion of the wavefunction, and $u_{\textbf{k}}(\textbf{r})$ is the periodic portion of the Bloch wavefunction and is the solution to the following equation:
\begin{align}
  \left( -\frac{(\nabla+i\textbf{k})^2}{2} + V(\textbf{r})\right) u_\textbf{k}(\textbf{r}) 
  = \varepsilon_\textbf{k}^0 u_\textbf{k}(\textbf{r}).
  \label{eq_nr_kohnsham}
\end{align}
By inserting the Bloch wavefunction in Eq.~(\ref{eq_bloch_wavefunction}), the SOC portion of the orbital energy  expressed in Eq.~(\ref{eq_soc_eigenenergy}) becomes 
\begin{align}
  \varepsilon_{\textbf{k}\pm}^{\textrm{SOC}} &= \textbf{s}^{\pm}_\textbf{k} \cdot 2
  \int d^3 r \nabla V({\textbf{r}}) \times\{
  \textbf{k} |u_{\textbf{k}}(\textbf{r})|^2
  - i u^*_{\textbf{k}}(\textbf{r})\nabla u_{\textbf{k}}(\textbf{r})
  \}
  \label{eq_e_k1},
\end{align}
where each Bloch state has a $\textbf{k}$-dependent spin moment expressed as follows:
\begin{align}
  \textbf{s}^{\pm}_\textbf{k} &\equiv \sum_{\sigma\sigma'} 
  \chi^{\pm*}_{\textbf{k}\sigma}\textbf{s}_{\sigma\sigma'} \chi^{\pm}_{\textbf{k}\sigma'}.
\end{align}
Equation~(\ref{eq_e_k1}) indicates a Zeeman-like term, which causes energy splitting between two spin states that are parallel and anti-parallel to the intrinsic magnetic field.
To analyze the $\textbf{k}$-dependence of the SOC energy in the vicinity of $\textbf{k} = \textbf{0}$, we expand the periodic portion of the wavefunction $u_{\textbf{k}}$ as follows:
\begin{align}
  u_{\textbf{k}}(\textbf{r}) &= u_{\textbf{0}}(\textbf{r}) 
  + \sum_{\beta = \{x,y\}} k_\beta \left. \frac{\partial u_{\textbf{k}}(\textbf{r})}{\partial k_{\beta}} \right|_{\textbf{k}=\textbf{0}}
  \nonumber \\
  &+ \frac{1}{2}\sum_{\beta,\beta' = \{x,y\}}  k_\beta k_{\beta'}
  \left.\frac{\partial^2 u_{\textbf{k}}(\textbf{r})}{\partial k_{\beta}\partial k_{\beta'}}\right|_{\textbf{k}=\textbf{0}}
  +O(k^3),
  \label{eq_u_expand}
\end{align}
where the first and second derivatives of the wavefunction with respect to $k$ are obtained by solving the Sternheimer Eqs.~\cite{PhysRev.96.951}
\begin{align}
  \left( -\frac{\nabla^2}{2}+V(\textbf{r}) - \varepsilon_\textbf{0}^0\right)
  \left.\frac{\partial u_\textbf{k}(\textbf{r})}{\partial k_{\beta}} \right|_{\textbf{k}=\textbf{0}}
  = i \frac{\partial u_\textbf{0}(\textbf{r})}{\partial r_\beta},
  \label{eq_u_first}
\end{align}
and
\begin{widetext}
\begin{align}
  \left( -\frac{\nabla^2}{2}+V(\textbf{r}) - \varepsilon_\textbf{0}^0\right)
  \left.\frac{\partial^2 u_{\textbf{k}}(\textbf{r})}{\partial k_{\beta}\partial k_{\beta'}}\right|_{\textbf{k}=\textbf{0}} &=
  \left(\left. \frac{\partial^2 \varepsilon_\textbf{k}^0}{\partial k_{\beta}\partial k_{\beta'}} \right|_{\textbf{k}=\textbf{0}}
  +\delta_{\beta \beta'}\right)u_{\textbf{0}}(\textbf{r})
  +i \left(\frac{\partial}{\partial r_\beta} \left.\frac{\partial u_{\textbf{k}}(\textbf{r})}{\partial k_{\beta'}} \right|_{\textbf{k}=\textbf{0}}
  +\frac{\partial}{\partial r_{\beta'}} \left.\frac{\partial u_{\textbf{k}}(\textbf{r})}{\partial k_{\beta}} \right|_{\textbf{k}=\textbf{0}}
  \right),
\end{align}
\end{widetext}
respectively.
As $u_{\textbf{0}}$ is the solution of the eigen-equation presented in Eq.~(\ref{eq_nr_kohnsham}) at $\textbf{k}=\textbf{0}$, we can set $u_{\textbf{0}}$ as a real function.
Consequently, $\partial u_\textbf{k}/\partial k_{\beta}|_{\textbf{k}=\textbf{0}}$ and $\partial^2 u_\textbf{k}/(\partial k_{\beta} \partial k_{\beta'})|_{\textbf{k}=\textbf{0}}$ become a pure imaginary and a real function, respectively.
Therefore, the complex conjugate of the wavefunction becomes
\begin{align}
  u^*_{\textbf{k}}(\textbf{r}) &= u_{\textbf{0}}(\textbf{r}) 
  - \sum_{\beta = \{x,y\}} k_\beta \left. \frac{\partial u_{\textbf{k}}(\textbf{r})}{\partial k_{\beta}} \right|_{\textbf{k}=\textbf{0}}
  \nonumber \\
  &+ \frac{1}{2}\sum_{\beta,\beta' = \{x,y\}}  k_\beta k_{\beta'}
  \left.\frac{\partial^2 u_{\textbf{k}}(\textbf{r})}{\partial k_{\beta}\partial k_{\beta'}}\right|_{\textbf{k}=\textbf{0}}
  +O(k^3).
  \label{eq_us_expand}
\end{align}
By inserting Eqs.~(\ref{eq_u_expand}) and (\ref{eq_us_expand}) into the SOC-energy term in Eq.~(\ref{eq_e_k1}) and applying partial integration and antisymmetry to the outer product, the zeroth- and second-order terms for $k$ vanish, and the energy is written as
\begin{widetext}
\begin{align}
  \varepsilon_{\textbf{k}\pm}^{\textrm{SOC}} = \textbf{s}^{\pm}_\textbf{k} \cdot 2
  \int d^3 r \nabla V({\textbf{r}}) \times \left\{
  \textbf{k} u^2_{\textbf{0}}(\textbf{r})
  +2 i \sum_{\beta=\{x,y\}} k_\beta \left.\frac{\partial u_{\textbf{k}}(\textbf{r})}{\partial k_\beta}\right|_{\textbf{k}=\textbf{0}}\nabla u_{\textbf{0}}(\textbf{r})
  \right\} + O(k^3).
  \label{eq_e_ksoc}
\end{align}
\end{widetext}
The dominant region of this integral is near the nuclear position where the absolute value of $\nabla V$ becomes large.
If $u_\textbf{0}$ comprises orbitals such as $s$, $p_z$, and $d_{3z^2-r^2}$, which are symmetric in the $x-y$ plane, then $\partial u_\textbf{0}/\partial r_{x(y)}$ is zero at the nuclear position.
Based on Eq.~(\ref{eq_u_first}), $\partial u_\textbf{0}/\partial k_{x(y)}$ becomes small in that region, and the integral in Eq.~(\ref{eq_e_ksoc}) is dominated by the first term.
In Sec.~\ref{sec_example}, we show that this orbital symmetry is realized in the Ag/Au QWS system.
\begin{figure}[!tb]
  \begin{center}
    \includegraphics[width=8.6cm]{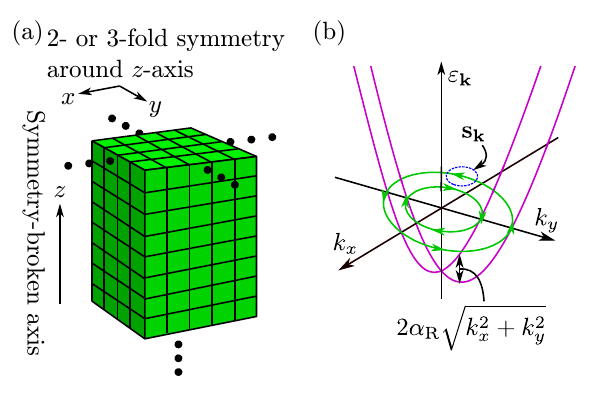}
    \caption{\label{fig_surface}
    (a) Schematic illustration of a surface system and (b) band structure.
    The system exhibits two- or three-fold symmetry on the $x-y$ plane, whereas reversal symmetry is broken  on the $z$-axis.
    The solid green lines indicate the equienergy line, and the arrows indicate the spin moment $\textbf{s}_\textbf{k}$.
    }
  \end{center}
\end{figure}
If we consider a system that exhibits two- or three-fold symmetry on the $x-y$ plane and no inversion symmetry on the $z$-axis [See Fig.~\ref{fig_surface}(a)], then only the $z$-component of $\nabla V$ remains in the integral and Eq.~(\ref{eq_e_ksoc}) becomes
\begin{align}
  \varepsilon_{\textbf{k}\pm}^{\textrm{SOC}} \approx 2 (\textbf{e}_z\times\textbf{k})\cdot\textbf{s}^{\pm}_\textbf{k}
  \int d^3 r \frac{\partial V(\textbf{r})}{\partial z} u^2_\textbf{0}(\textbf{r})
\end{align}
in the vicinity of $\textbf{k} = \textbf{0}$.
The energy difference between two-spin states depicted in Fig.~\ref{fig_surface}(b) is expressed as follows:
\begin{align}
  \left|\varepsilon_{\textbf{k}+}^{\textrm{SOC}} - \varepsilon_{\textbf{k}-}^{\textrm{SOC}} \right| &= 2 \alpha_\textbf{R} \sqrt{k_x^2+k_y^2},
\end{align}
where we use $|\textbf{s}_\textbf{k}|=1/2$ and define the following Rashba-splitting parameter
\begin{align}
  \alpha_\textrm{R} &\equiv \int d^3 r \frac{\partial V(\textbf{r})}{\partial z} u^2_\textbf{0}(\textbf{r}).
  \label{eq_rashba1}
\end{align}
The applicability of Eq.~(\ref{eq_rashba1}) is not limited to the two-dimensional free electronic system, in contrast to an earlier study~\cite{Nagano_2009}, even though the same formulation is derived.
Next, we discuss the Rashba-splitting parameter $\alpha_\textrm{R}$.

\subsection{QWS}

\begin{figure}[!tb]
  \begin{center}
    \includegraphics[width=8.6cm]{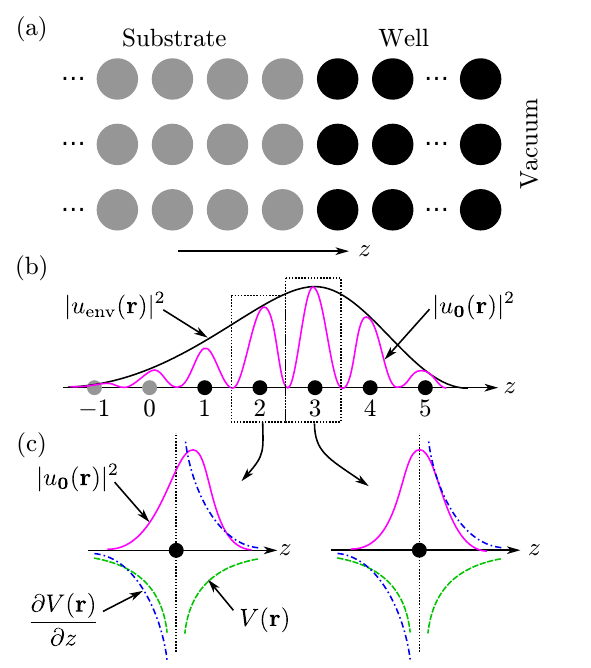}
    \caption{\label{fig_envelope}
    (a) Schematic illustration of a system comprising a semi-infinite substrate and a finite-layered well region. 
    Thw black (gray) circles indicate the well (substrate) atoms.
    The system is periodic for upward and downward directions.
    (b) Schematic illustration of QWS.
    The solid magenta and black lines indicate the square of the wavefunction and its envelope function, respectively.
    The index of sites begins from zero at the substrate site at the interface and increases (decreases) in the well (substrate) direction.
    (c) Schematic illustration of behavior of the integrant in Eq.~(\ref{eq_rashba1}) at sites 2 and 3.
    }
  \end{center}
\end{figure}
Let us consider the schematic illustration of the QWS shown in Fig.~\ref{fig_envelope}(a).
If the considered electronic state is energetically isolated, then the wavefunction can be approximated as single-band state smoothly modulated along the $z$-axis, i.e., $u_\textbf{0}(\textbf{r})\approx \varphi_{\textrm{at}}(\textbf{r}-\textbf{r}_i)u_{\textrm{env}}(z)$, in the vicinity of the $i$-th nucleus, where $\varphi_{\textrm{at}}$ is the atomic orbital and $u_{\textrm{env}}$ is the envelope function [See Fig.~\ref{fig_envelope}(b)].
Because $V(\textbf{r})$ has a large gradient in the vicinity of the nuclei, the integrand of Eq.~(\ref{eq_rashba1}) can be depicted schematically as shown in Fig.~\ref{fig_envelope}(c).
Subsequently, the Rashba parameter becomes
\begin{align}
  \alpha_{\textrm{R}}
  &\approx
  \sum_{i = -\infty}^{N_{\textrm{well}}} \int d^3 r \frac{\partial V_{\textrm{at}}(\textbf{r}-\textbf{r}_i)}{\partial z}
  |\varphi_{\textrm{at}}(\textbf{r}-\textbf{r}_i)|^2|u_{\textrm{env}}(z)|^2
  \nonumber \\
  &\equiv
  \sum_{i = -\infty}^{N_{\textrm{well}}} \Delta \alpha_{\textrm{R}, i},
  \label{eq_rashba2}
\end{align}
where $V_{\textrm{at}}$ and $\textbf{r}_i$ are the potential and position of the $i$-th nucleus, respectively;
$N_{\textrm{well}}$ is the number of monolayers in the well region.
The relatively smooth envelope function can be approximated around the $i$-th nucleus as $|u_{\textrm{env}}(z)|^2=|u_{\textrm{env}}(z_i)|^2+d |u_{\textrm{env}}|^2/dz|_{z=z_i}(z-z_i)$. Thus, we have
\begin{align}
  \Delta \alpha_{\textrm{R}, i}
  &\approx
  |u_{\textrm{env}}(z_i)|^2
  \int d^3 r \frac{\partial V_{\textrm{at}}(\textbf{r})}{\partial z} |\varphi_{\textrm{at}}(\textbf{r})|^2
  \nonumber \\
  &+
  \left. \frac{d |u_{\textrm{env}}|^2}{d z}\right|_{z=z_i}
  \int d^3 r \frac{\partial V_{\textrm{at}}(\textbf{r})}{\partial z} |\varphi_{\textrm{at}}(\textbf{r})|^2 z
  \nonumber \\
  &=\left. \frac{d |u_{\textrm{env}}|^2}{d z}\right|_{z=z_i}
  \langle \delta V\rangle_i,
  \label{eq_rashba3}
\end{align}
where 
\begin{align}
  \langle \delta V\rangle_i \equiv
  \int d^3 r \frac{\partial V_{\textrm{at}}(\textbf{r})}{\partial z} |\varphi_{\textrm{at}}(\textbf{r})|^2 z.
\end{align}
The first term on the first line of Eq.~(\ref{eq_rashba3}) vanishes because of the symmetries of $V_{\textrm{at}}$ and $\varphi_{\textrm{at}}$.
$\Delta \alpha_{\textrm{R}, i}=0$ when the gradient of the envelope function is 0 [right panel of Fig.~\ref{fig_envelope}(c)], whereas $\Delta \alpha_{\textrm{R}, i}$ is positive (negative) when the envelope function increases (decreases) as shown in the left panel of Fig.~\ref{fig_envelope}~(c).
Substituting Eq.~(\ref{eq_rashba3}) into Eq.~(\ref{eq_rashba2}) yields the total Rashba parameter as follows:
\begin{align}
  \alpha_\textrm{R} = \sum_{i = -\infty}^{N_{\textrm{well}}} \left. \frac{d |u_{\textrm{env}}|^2}{d z}\right|_{z=z_i}
  \langle \delta V\rangle_i.
\end{align}
In the structure shown in Fig.~\ref{fig_envelope}~(b), because the averaged gradient of the potential $\langle \delta V \rangle_i$ in this summation assumes only two values, i.e., the averages for the substrate ($\langle\delta V\rangle_{\textrm{sub}}$) and well ($\langle\delta V\rangle_{\textrm{well}}$), we can shift the averaged gradients of the potential outside $\sum$ as follows:
\begin{align}
  \alpha_\textrm{R}
  &\approx
  \langle\delta V\rangle_{\textrm{sub}}
  \sum_{i = -\infty}^{0}
  \left. \frac{d |u_{\textrm{env}}(z)|^2}{d z} \right|_{z=z_i}
  \nonumber \\
  &+
  \langle\delta V\rangle_{\textrm{well}}
  \sum_{i = 1}^{N_{\textrm{well}}}
  \left. \frac{d |u_{\textrm{env}}(z)|^2}{d z} \right|_{z=z_i}.
\end{align}
If the envelope function $|u_{\textrm{env}}(z)|^2$ varies gradually, then these summations can be approximated with integrals as
\begin{align}
  \alpha_\textrm{R}
  &\approx
  \frac{\langle\delta V\rangle_{\textrm{sub}}}{c_{\textrm{sub}}}
  \int_{-\infty}^{z_\textrm{b}} dz
  \frac{d |u_{\textrm{env}}(z)|^2}{d z}
  \nonumber \\
  &+
  \frac{\langle\delta V\rangle_{\textrm{well}}}{c_{\textrm{well}}}
  \int_{z_\textrm{b}}^{z_\textrm{b}+c_{\textrm{well}}N_{\textrm{well}}} dz
  \frac{d |u_{\textrm{env}}(z)|^2}{d z}
  \nonumber \\
  &=
  \left(
    \frac{\langle\delta V\rangle_{\textrm{sub}}}{c_{\textrm{sub}}}
  -
  \frac{\langle\delta V\rangle_{\textrm{well}}}{c_{\textrm{well}}}
  \right)
  |u_{\textrm{env}}(z_\textrm{b})|^2,
  \label{eq_rashba4}
\end{align}
where $c_{\textrm{sub}}$ ($c_{\textrm{well}}$) is the layer distance at the substrate (well) and $z_\textrm{b}$ is the boundary of the substrate and well.
Furthermore, we leverage the fact that the envelope function vanishes in the deep position of the substrate ($z=-\infty$) and on the surface ($z=z_\textrm{b}+c_{\textrm{well}}N_{\textrm{well}}$).
This formula shows that $\alpha_\textrm{R}$ is determined by the charge density at the interface regardless of the detailed structure inside the well, provided that the atomic species used for the film is known.
Additionally, the difference in the potential gradients $\partial V / \partial z$ of the two atoms significantly enhances the Rashba effect.

\subsection{Tight-binding model for QWS and analytical solution}

\begin{figure}[!tb]
  \begin{center}
    \includegraphics[width=8.6cm]{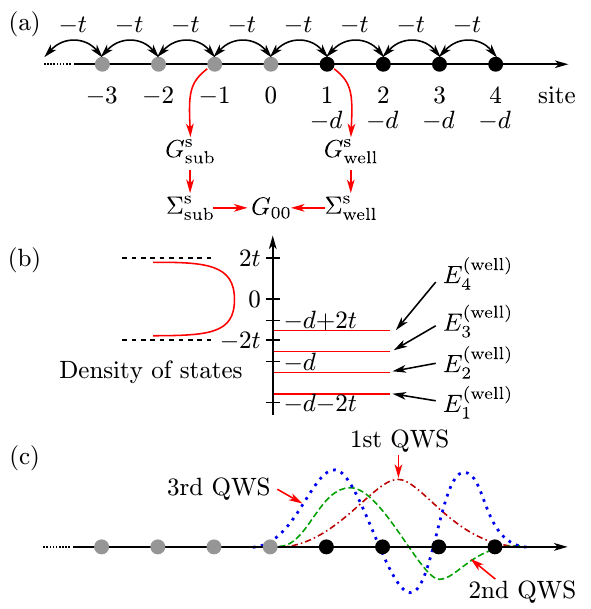}
    \caption{\label{fig_tightbinding}
    (a) One-dimensional tight-binding model for QWS.
    (b) Schematic illustration of density of states for isolated semi-infinite substrate (left) and finite-length well region (right).
    If these separated systems hybridize, then localized states are formed in the well region (c).
    In this case, the fourth QWS originating from the state with the energy of $E_4^{(\textrm{well})}$ is absent because that state deeply hybridizes the continuum of the substrate states.
    Dash-dotted red, dashed green, and dotted blue lines indicate the first, second, and third QWS, respectively.
    }
  \end{center}
\end{figure}
In this section, we estimate the $|u_{\textrm{env}}(z_\textrm{b})|^2$ of a QWS as the wavefunction of the simplified one-dimensional tight-binding model shown in Fig.~\ref{fig_tightbinding}(a).
Subsequently, the envelope function at each atomic site $u_{\textrm{env}}(z_i)$ is approximated as a coefficient for the local basis of the tight-binding model as
\begin{align}
  u_{\textrm{env}}(z_i) \approx \langle i | u_n \rangle,
\end{align}
where $|i\rangle$ is the local basis at site $i$ and $| u_n \rangle$ is the $n$-th QWS.
We focus on the square of this quantity at the interfacial substrate site, namely $i=0$, and compute this quantity as a residue of Green's function at a pole of a QWS as
\begin{align}
  |u_{\textrm{env}}(z_\textrm{b})|^2 \approx |\langle 0 |u_n\rangle|^2 = \textrm{Res}_{\varepsilon=\varepsilon_n}[ G_{00}(\varepsilon)],
  \label{eq_uenv_res}
\end{align}
where the pole $\varepsilon_n$ corresponds to the eigenenergy of the $n$-th QWS.
Subsequently, Green's function of the interfacial substrate can be written as a single-site Dyson's equation [see Fig.~\ref{fig_tightbinding}(a)]
\begin{align}
  G_{00}(\varepsilon) = \frac{1}{\varepsilon-\Sigma^\textrm{s}_{\textrm{sub}}-\Sigma^\textrm{s}_{\textrm{well}}},
  \label{eq_green}
\end{align}
where
\begin{align}
  \Sigma^\textrm{s}_{\textrm{sub}}(\varepsilon) = t^2 G_{\textrm{sub}}^\textrm{s}(\varepsilon)
\end{align}
and
\begin{align}
  \Sigma^\textrm{s}_{\textrm{well}}(\varepsilon) = t^2 G_{\textrm{well}}^\textrm{s}(\varepsilon)
\end{align}
are the local self-energies originating from the two neighboring surfaces.
$G_{\textrm{sub}}^\textrm{s}$ and $G_{\textrm{well}}^\textrm{s}$ are the surface Green's functions at the substrate and well, respectively.
$G_{\textrm{sub}}^\textrm{s}$ is computed using the recursion method~\cite{RHaydock_1975} as an onsite edge Green's function of semi-infinite chain
\begin{align}
  G_{\textrm{sub}}^\textrm{s}(\varepsilon) &= \frac{\varepsilon+\sqrt{\varepsilon^2-4t^2}}{2t^2}.
\end{align}
Meanwhile, $G_{\textrm{well}}^\textrm{s}$ becomes
\begin{align}
  G_{\textrm{well}}^\textrm{s}(\varepsilon) &= \sum_{n} 
  \frac{\left\langle 1 \left|\phi_n^{(\textrm{well})}\right.\right\rangle\left\langle \left.\phi_n^{(\textrm{well})} \right| 1\right\rangle}
  {\varepsilon-E_n^{(\textrm{well})}}.
  \label{eq_green_ag}
\end{align}
Here, $\left|\phi_n^{(\textrm{well})}\right\rangle$ and $E_n^{(\textrm{well})}$ are eigenvectors and eigenvalues for a finite-length chain whose Hamiltonian is expressed as
\begin{align}
  \hat{H}_0 =
  \begin{pmatrix}    
    -d & -t &    &         &    &    \\
    -t & -d & -t &         &    &    \\
       & -t & -d & -t      &    &    \\
       &    & -t & \ddots\ & -t &    \\
       &    &    & -t      & -d & -t \\
       &    &    &         & -t & -d
  \end{pmatrix},
\end{align}
where the size of this matrix is the number of the well layers $N_{\textrm{well}}$.
This matrix is known as the tridiagonal Toeplitz matrix, which can be diagonalized analytically~\cite{https://doi.org/10.1002/nla.1811,KULKARNI199963} as
\begin{align}
  E_n^{(\textrm{well})} &= -d - 2 t \cos(n\omega),
  \\
  \left|\phi_n^{(\textrm{well})}\right\rangle &= \sqrt{\frac{2}{N_{\textrm{well}}+1}}
  \nonumber \\
  &\times\left(\sin(n\omega), \sin(2n\omega), \cdots, \sin(nN_{\textrm{well}}\omega) \right)^T,
  \\
  \omega &= \frac{\pi}{N_{\textrm{well}}+1}.
\end{align}
Next, we focus on the $n$-th QWS, whose energy is close to $E_n^{(\textrm{well})}$.
To stabilize this state, the following conditions must be satisfied:
\begin{align}
  E_n^{(\textrm{well})} < -2 t, \quad N_{\textrm{well}} \geq n.
\end{align}
When we focus on the energy $E_n^{(\textrm{well})}$, we are only required to consider the corresponding contribution in Eq.~(\ref{eq_green_ag}).
Subsequently, for $\varepsilon=E_n^{(\textrm{well})}+i\gamma$, we have
\begin{align}
  \Sigma^\textrm{s}_{\textrm{well}}(\varepsilon) \approx \frac{t^2}{\varepsilon-E_n^{(\textrm{well})}}\frac{2}{N_{\textrm{well}}+1} \sin^2(n\omega).
  \label{eq_sigma_ag}
\end{align}
Next, we consider $\Sigma_{\textrm{sub}}(\varepsilon)$ expressed as
\begin{align}
  \Sigma^\textrm{s}_{\textrm{sub}}(\varepsilon) = t^2 
  \left(\frac{\varepsilon+\sqrt{\varepsilon^2-4t^2}}{2t^2}\right)
\end{align}
By performing the Taylor expansion around $\varepsilon-E_n^{(\textrm{well})}$, we have
\begin{align}
  \Sigma^\textrm{s}_{\textrm{sub}}(\varepsilon) &= 
  \frac{1}{2} \left(E_n^{(\textrm{well})} + \sqrt{E_n^{(\textrm{well})2}-4t^2}\right)
  \nonumber \\
  &+\left(\frac{1}{2}+\frac{E_n^{(\textrm{well})}}{2\sqrt{E_n^{(\textrm{well})2}-4t^2}}\right)
  (\varepsilon-E_n^{(\textrm{well})})
  \nonumber \\
  &+ O\left[\left(\varepsilon-E_n^{(\textrm{well})}\right)^2\right]
  \label{eq_sigma_au}
\end{align}
Using Eqs.~(\ref{eq_sigma_ag}) and (\ref{eq_sigma_au}), $G(\varepsilon)$ in Eq.~(\ref{eq_green}) can be approximated as
\begin{align}
  G(\varepsilon) &\approx 
  \frac{1}{\varepsilon-\frac{e_2}{\varepsilon-E_n^{(\textrm{well})}}+e_1+e_0 (\varepsilon-E_n^{(\textrm{well})})},
  \label{eq_green_approx}
  \\
  e_0 &\equiv \frac{1}{2}+\frac{E_n^{(\textrm{well})}}{2\sqrt{E_n^{(\textrm{well})2}-4t^2}},
  \\
  e_1 &\equiv \frac{1}{2}\left(E_n^{(\textrm{well})} + \sqrt{E_n^{(\textrm{well})2}-4t^2}\right),
  \\
  e_2 &\equiv \frac{2 t^2}{N_{\textrm{well}}+1} \sin^2(n\omega).
\end{align}
Based on this formula of Green's function, we can compute the pole associated to with the QWS as follows:
\begin{align}
  \varepsilon_n &= E_n^{(\textrm{well})} 
  \nonumber \\ 
  &+ \frac{e_1-E_n^{(\textrm{well})}\pm\sqrt{4e_2(1-e_0)+(e_1-E_n^{(\textrm{well})})^2}}
  {2(1-e_0)}.
\end{align}
The minus sign is relevant to satisfy the condition $\varepsilon_n = E_n^{(\textrm{well})}$ when $t=0$ and $d>0$.
The corresponding residue is
\begin{align}
  &\textrm{Res}_{\varepsilon=\varepsilon_n}[ G_{00}(\varepsilon)] 
  \nonumber \\
  &= \frac{\sqrt{4e_2(1-e_0)+(e_1-E_n^{(\textrm{well})})^2}-(e_1-E_n^{(\textrm{well})})}
  {2 (1-e_0)\sqrt{4e_2(1-e_0)+(e_1-E_n^{(\textrm{well})})^2}}.
  \label{eq_residual}
\end{align}
By combining Eqs.~(\ref{eq_rashba4}) and (\ref{eq_uenv_res}), we obtain
\begin{align}
  \alpha_\textrm{R} \approx \alpha_\textrm{R}^{\delta V} \textrm{Res}_{\varepsilon=\varepsilon_n}[ G_{00}(\varepsilon)],
  \label{eq_alpha_residual}
\end{align}
where 
\begin{align}
  \alpha_\textrm{R}^{\delta V}\equiv
  \frac{\langle\delta V\rangle_{\textrm{sub}}}{c_{\textrm{sub}}}-
  \frac{\langle\delta V\rangle_{\textrm{well}}}{c_{\textrm{well}}}
  \label{eq_alpha_v}
\end{align}
governs the strength of the SOC in the Rashba effect, whereas $\textrm{Res}_{\varepsilon=\varepsilon_n}[ G_{00}(\varepsilon)]$ governs the shape of the electronic state.
Based on the discussion above, the only free parameter in Eq.~(\ref{eq_residual}) is the ratio between the onsite potential $d$ and the hopping integral $t$.
Therefore, in this analysis, only two parameters ($\alpha_\textrm{R}^{\delta V}$ and $d/t$) govern the Rashba splitting.

\section{Example in Ag/Au(111)} \label{sec_example}

\begin{figure}[!tb]
  \begin{center}
    \includegraphics[width=8.6cm]{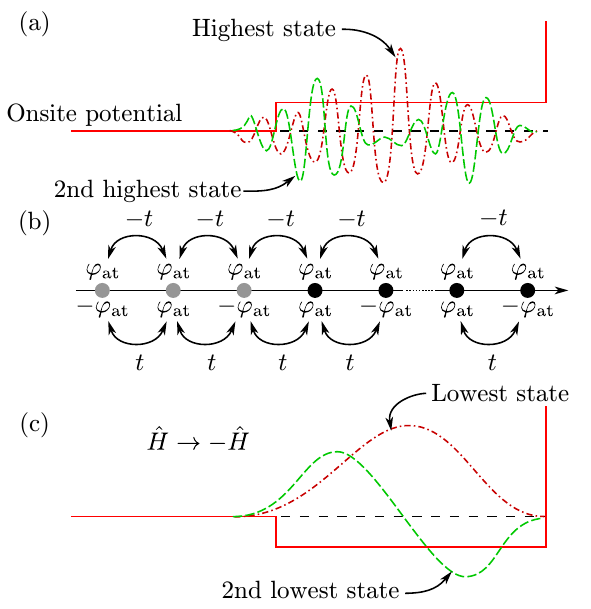}
    \caption{\label{fig_inverse}
    (a) Schematic illustration of QWS at the repulsive onsite potential.
    The solid, dashed-dotted, and dashed lines indicate onsite potential, highest QWS, and second-highest QWS, respectively.
    (b) Tight-binding model for the considered system. 
    The black and gray circles indicate sites in the quantum well and substrate.
    Changing the basis function from the ordinary (above the arrow) to the staggard (below the arrow) results in the opposite sign for the hopping parameter ($t$).
    Finally, changing the sign of the Hamiltonian ($\hat{H}\rightarrow-\hat{H}$) yields the QWS at an attractive onsite potential (c).
    }
  \end{center}
\end{figure}
In this section, as an example, we show the Rashba splitting in a few Ag MLs on an Au(111) surface.
Although the onsite potential of Ag is larger than the potential of Au ($d<0$)~\cite{PhysRevB.84.075412,CHIANG2000181}, the discussion in the previous section remains valid.
In this section, some modifications are introduced as follows:
The model Hamiltonian of the Ag/Au(111) system is
\begin{align}
  \hat{H} =
  \begin{pmatrix}    
    \ddots & \ddots &     &     &        &     \\
    \ddots & 0      & -t  &     &        &     \\
           & -t     & 0   & -t  &        &     \\
           &        & -t  & |d| & -t     &     \\
           &        &     & -t  & \ddots & -t  \\
           &        &     &     & -t     & |d|
  \end{pmatrix}.
\end{align}
Here, a positive onsite potential exists at the well region [see Fig.~\ref{fig_inverse}(a)].
If we change the sign of local orbitals staggered as in Fig.~\ref{fig_inverse}(b), then the sign of the hopping parameter $t$ is inverted as
\begin{align}
  \hat{H}' =
  \begin{pmatrix}    
    \ddots & \ddots &   &      &        &   \\
    \ddots & 0      & t &      &        &   \\
           & t      & 0 & t    &        &   \\
           &        & t & |d| & t      &   \\
           &        &   & t    & \ddots & t \\
           &        &   &      & t      & |d|
  \end{pmatrix}.
\end{align}
However, the eigenvalues remain unchanged.
By changing the sign as $\hat{H}'' = -\hat{H}'$, we obtain the Hamiltonian $\hat{H}''$ for the QWS system with an attractive onsite well potential [Fig.~\ref{fig_inverse}(c)].
Because the energy spectrums for $\hat{H}''$ and $\hat{H}$ are opposite, the QWSs appear in the highest energy states for a repulsive onsite well potential system, in contrast to the attractive potential system whose QWSs appear in the lowest energy states.

Next, we explain the numerical condition for the first-principles calculation.
The slab models of Ag/Au(111) were constructed as follows:
We used 60 MLs comprising $1\times1$ Au units to describe the substrate.
Above this Au substrate, we placed 6 -- 34 MLs comprising $1\times1$ Ag units and relaxed the interlayer distances.
To perform first-principles calculations based on DFT, we used an open-source program package, OpenMX~\cite{openmxsquare}, which employs norm-conserving pseudopotentials~\cite{PhysRevB.47.6728} and the linear combination of pseudo-atomic orbitals (LCPAO)~\cite{PhysRevB.67.155108,PhysRevB.69.195113} to solve the Kohn--Sham equation.
We utilized the exchange-correlation energy proposed by Perdew--Burke--Ernzerhof~\cite{PhysRevLett.77.3865} via the generalized gradient approximation.
Subsequently, we performed Brillouin-zone integrations on $14\times14\times1$ $\textbf{k}$-point grids and used the Fermi--Dirac distribution function at 5,000 K to smear the rapidly varying electronic distribution.
We treated the SOC by solving the Kohn--Sham equation in the spinor representation.
The basis set was configured as follows: Two, two, two, and one radial wavefunctions with a cutoff radius of 7.0 a.u. were used for the s-, p-, d-, and f-orbitals of the Ag atom (Ag7.0-s2p2d2f1), respectively, and four, three, two, and one radial wavefunctions with a cutoff radius of 7.0 a.u. for the corresponding orbitals of the Au atom (Au7.0-s4p3d2f1).
We placed an empty atom to precisely describe the electronic states in the vicinity of the surface.

\begin{figure}[!tb]
  \begin{center}
    \includegraphics[width=8.6cm]{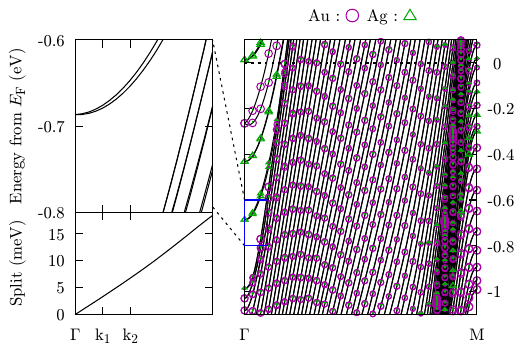}
    \caption{\label{fig_band}
    Band structure of Ag 24 MLs on Au(111).
    (Right) Band structure between $\Gamma$ and M points.
    The size of magenta circles and green triangles indicate the strength of Au and Ag atomic characteristics of each Kohn--Sham orbital, respectively.
    (Left top) Magnified figure showing the band structure of the second QWS in the vicinity of $\Gamma$ point.
    (Left bottom) The energy split of the band in the upper panel.
    Two $\textbf{k}$ points, $k_1$ and $k_2$, are the same as those shown in Fig.~\ref{fig_orb}.
    }
  \end{center}
\end{figure}
\begin{figure}[!tb]
  \begin{center}
    \includegraphics[width=8.6cm]{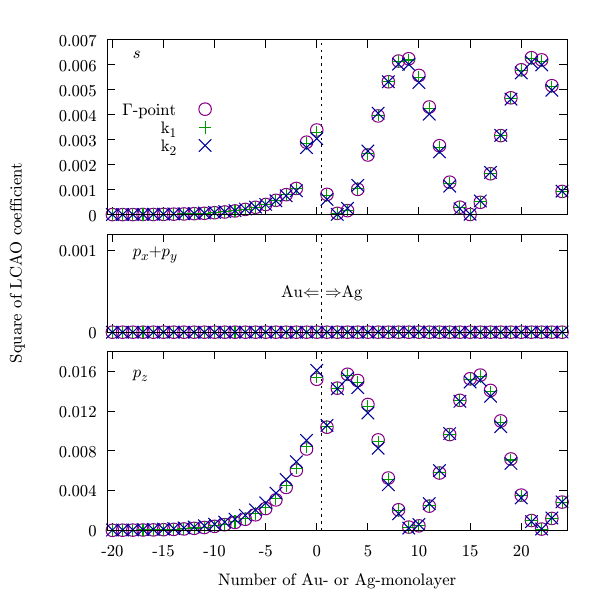}
    \caption{\label{fig_orb}
    Square of LCPAO coefficient for $s$ (top), $p_x+p_y$ (middle), and $p_z$ orbitals (bottom), for the band depicted at three $\textbf{k}$ points, i.e., $\Gamma$, $k_1$, and $k_2$, in the left panel of Fig.~(\ref{fig_band}).
    }
  \end{center}
\end{figure}
As a typical condition of this layered system, we present the band structure of 24 Ag MLs on Au(111) in Fig.~\ref{fig_band}.
The energy bands immediately below the Fermi level primarily comprise Au orbitals, whereas the isolated bands comprise Ag orbitals only near the $\Gamma$ point.
Additionally, near the $\Gamma$ point, the band slightly above the Fermi level is the surface state of the outermost Ag layer, whereas the next band is the surface state at the opposite Au surface in the slab model.
The next two bands are the first and second QWSs, respectively.
These bands are well separated from the other bands, and the Rashba energy splitting is almost $\textbf{k}$-linear around the $\Gamma$ point, as shown in the left panel of Fig.~\ref{fig_band}.
To present the detailed orbital characteristics of this isolated QWS, we show the LCPAO coefficient and its variation with respect to $\textbf{k}$ in Fig.~\ref{fig_orb}.
The depicted band and $\textbf{k}$-point are the same as those shown in the left panel of Fig.~\ref{fig_band}.
The dominant and subdominant contributions are from the $p_z$- and $s$-orbitals, respectively, whereas $p_x$ and $p_y$ contributed minimally to this electronic state.
This trend is the same as that observed in an earlier study for an Ag(111) surface~\cite{PhysRevB.90.235422}.
The $\textbf{k}$-dependence was weak in the region between $\Gamma$ and $k_2$ points.
The mechanism of the formation of this electronic state can be explained as follows:
\begin{figure}[!tb]
  \begin{center}
    \includegraphics[width=8cm]{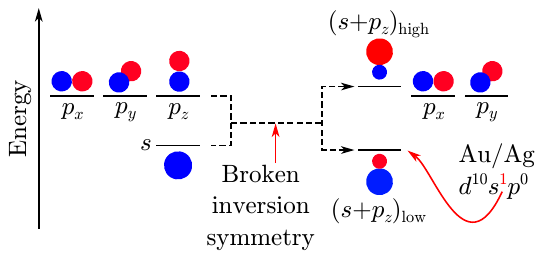}
    \caption{\label{fig_spz}
    Schematic illustration of the energy diagram of the Au/Ag system.
    Symmetry breaking along $z$-direction primarily affects $s$- and $p_z$-orbitals, which hybridize.
    Valence top electron primarily occupies lower $s+p_z$ state.
    }
  \end{center}
\end{figure}
Ag (Au) fully occupies the 4$d$ (5$d$) states and one additional electron occupies the $s$ and $p$ states.
The symmetry breaking along the $z$-direction primarily affects the $s$- and $p_z$-orbitals, which hybridize, as shown in Fig.~\ref{fig_spz}.
Because the valence electronic configuration of Au and Ag is $d^{10}s^1p^0$, the valence top electron primarily occupies the lower $s+p_z$ state.
The subdominant $s$-orbital in Fig.~\ref{fig_orb} shifts the $p_z$-orbital, as shown in the right panel of Fig.~\ref{fig_spz};
this shift depends on the position because the peaks and dips for $s$-LCPAO differ from those of $p_z$-LCPAO in Fig.~\ref{fig_orb}.
This position-dependent shift results in the envelope function of the QWS, as shown in Fig.~\ref{fig_envelope}.

\begin{figure}[!tb]
  \begin{center}
    \includegraphics[width=8.6cm]{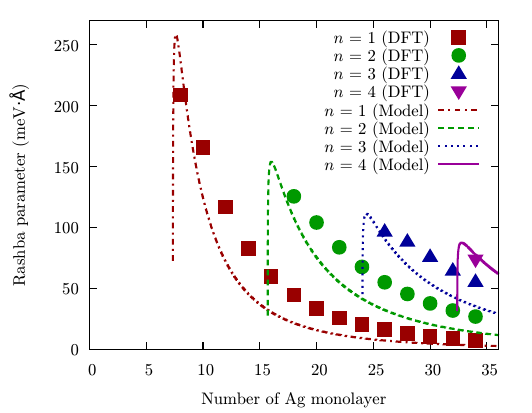}
    \caption{\label{fig_split}
    Rashba-splitting parameter of QWSs ($n=$ 1, 2, 3, and 4) in Ag/Au(111).
    Symbols indicate parameters computed via DFT, whereas lines indicate results obtained using Eq.~(\ref{eq_alpha_residual}) by setting $d/t=0.14$ and $\alpha_\textrm{R}^{\delta V}=47,000$ (meV$\cdot$\AA).
    }
  \end{center}
\end{figure}
Figure~\ref{fig_split} shows the Rashba-splitting parameter for the QWSs ($n=$ 1, 2, 3, and 4) in Ag/Au(111) based on first-principles calculations.
In the same figure, we present the result obtained using the formula expressing the tight-binding model [Eq.~(\ref{eq_alpha_residual})] by setting $d/t=0.14$ and $\alpha_\textrm{R}^{\delta V}=47,000$ (meV$\cdot$\AA).
As shown, we can qualitatively fit the Rashba splitting by adjusting only two parameters ($\alpha_\textrm{R}^{\delta V}$ and $d/t$).
Our analysis based on a simple one-dimensional tight-binding model can capture the Rashba energy-splitting behavior of the QWS.
Moreover, because the DFT result agrees quantitatively with the result from the ARPES experiment~\cite{PhysRevB.104.L180409}, this tight-binding model analysis closely reproduces the experimental result.

\begin{figure}[!tb]
  \begin{center}
    \includegraphics[width=8.6cm]{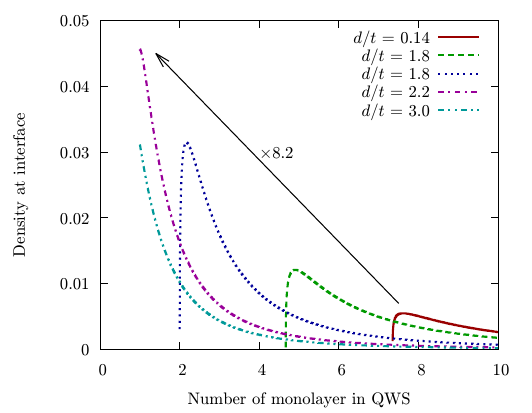}
    \caption{\label{fig_split2}
    Interfacial density obtained using Eq.~(\ref{eq_residual}) by varying $d/t$ and the number of ML while fixing $n=1$.
    }
  \end{center}
\end{figure}
Finally, we predicted the ideal condition based on our derived formula.
Figure \ref{fig_split2} shows the density of the first ($n=1$) QWS at the interface [Eq.~(\ref{eq_residual})] by varying the ratio $d/t$.
This function decays proportionally to $(N_{\textrm{well}}+1)^{-3}$ at the large-$N_{\textrm{well}}$ limit.
While the peak ML number of the interfacial density decreases, the peak height increases with the $d/t$ ratio.
Subsequently, after the peak position achieves $N_{\textrm{well}}=1$, the maximum value decreases as the $d/t$ ratio increases since $N_{\textrm{well}}$ is equal to or greater than 1.
This function achieves the maximum value when $d/t=2.2$ and $N_{\textrm{well}}=1$, and the value is 8.2 times larger than that obtained at $d/t=0.14$, which is a realistic setting (see Fig.~\ref{fig_split}).
Because the model yields a maximum value at $N_{\textrm{well}}=1$, the Bi (1 ML) on Cu(111)~\cite{PhysRevLett.104.066802} and the Bi surface alloy (1/3 ML) on Ag~\cite{PhysRevLett.98.186807} have large Rashba splitting parameters. 
This is the optimal condition for achieving large Rashba splitting in the QWS while disregarding the dependence on the SOC strength.

\section{Summary} \label{sec_summary}

We formulated a theory to obtain the $\textbf{k}$-linear Rashba splitting energy systematically from the SOC Hamiltonian by employing reasonable assumptions, such as broken inversion symmetry, two- or three-fold symmetry around the $z$-axis, and a steep nuclear potential.
Subsequently, we derived a minimum model that captures the Rashba effect in a QWS using a one-dimensional tight-binding model and Green's function.
Our theory agreed qualitatively with the first-principles result of a realistic Ag/Au(111) system using only two fitting parameters.
Furthermore, we predicted the optimal condition to obtain a large Rashba splitting.
This result can be utilized to design new material systems for realizing efficient spintronics devices.
Moreover, our theoretical framework can be extended to, for example, more complex nanostructures such as three-step QWSs.

\begin{acknowledgments}
  We acknowledge Ryo Noguchi, Kenta Kuroda, and Takeshi Kondo for the fruitful discussions.
  M.K. was supported by KAKENHI 20K15012 from MEXT, Japan.
  The numerical calculations were performed on the supercomputers at the Institute for Solid State Physics at the University of Tokyo.
\end{acknowledgments}


\bibliography{ref}

\end{document}